 \documentstyle[12pt]{article}

\newcommand\mathC{{\mkern1mu\raise2.2pt\hbox{$\scriptscriptstyle|$}
        {\mkern-7mu\rm C}}}
\newcommand{\mathR}{{\rm I\! R}}                

\renewcommand\[{[\,}                            
\renewcommand\]{\,]}                            

\renewcommand\mathR{{\rm I\! R}}

\newcommand\unit{{\rm 1\kern-3.2pt I}}

\let \ssection = \section
\renewcommand{\section}{\setcounter{equation}{0} \ssection}
\pagebreak

\begin{document}
\title{Poincar\'{e} invariance for continuous-time histories }
\author{Ntina Savvidou \thanks{ntina@ic.ac.uk}\\ {\small Theoretical
Physics Group, The Blackett Laboratory,} \\ {\small Imperial
College, SW7 2BZ, London, UK} \\ } \maketitle
\begin{abstract}
We show that the relativistic analogue of the two types of time translation in a
non-relativistic history theory is the existence of two distinct {\em Poincar\'{e}
groups\/}.  The `internal' Poincar\'{e} group is analogous to the one that arises in
the standard canonical quantisation scheme; the `external' Poincar\'{e} group is
similar to the group that arises in a {\em Lagrangian\/} description of the standard
theory. In particular, it {\em performs explicit changes of the spacetime foliation}
that is implicitly assumed in standard canonical field theory.
\end{abstract}

\pagebreak

\section{Introduction}
The generalisation of continuous-time history theory to include relativistic quantum
fields raises some subtle issues that tend to be hidden in the normal canonical
treatment of a quantum field.

The standard canonical quantisation of a relativistic field requires the choice of a
Lorentzian foliation on the background spacetime: the Hamiltonian is then defined with
respect to this foliation. There exist many unitarily inequivalent representations of
the canonical commutation relations for this quantum field theory: the physically
appropriate one is chosen by requiring that the Hamiltonian exists as a well-defined
self-adjoint operator. In this sense---like the Hamiltonian itself---the physically
appropriate representation is foliation-dependent. Relativistic covariance is then
implemented by seeking a representation of the Poincar\'{e} group on the resulting
Hilbert space. However, the Poincar\'{e} group thus constructed does not explicitly
perform a {\em change\/ of the foliation}.

The HPO continuous-time histories approach to quantum theory
\cite{Sav99,HPOHis,SA00,An00} is particularly suited to deal with systems that have a
non-trivial temporal structure, and therefore it should be able to provide a
significant clarification of this point.

Specifically, we will show that the relativistic analogue of the two types of time
translation that arise in a non-relativistic history theory is the existence of two
distinct {\em Poincar\'{e} groups\/}. The `internal' Poincar\'{e} group is analogous
to the one that arises in the standard canonical quantisation scheme as sketched
above.

However, the `external' one is a novel object: it is similar to the group that arises
in the  {\em Lagrangian\/} description of the field theory. In particular, it
explicitly performs {\em changes\/} of the foliation. This arises from the striking
property that HPO theories admit two distinct types of time transformation, each
representing a distinct quality of time \cite{Sav99}. The first corresponds to time
considered purely as a kinematical parameter of a physical system, with respect to
which a history is defined as a succession of possible events. It is strongly
connected with the temporal-logical structure of the theory and is related to the view
of time as a parameter that determines the ordering of events. The second corresponds
to the dynamical evolution generated by the Hamiltonian. For a detailed presentation
of the HPO continuous-time programme see \cite{Sav99}.

As we shall see, one of the important results of the formalism as applied to a field
theory is that, even though the representations of the history algebra are foliation
dependent, the physical quantities (probabilities) are {\em not}.

In section 2, we shall give a brief description of the underlying concepts of the
continuous-histories programme: this is necessary for establishing the framework of
the ensuing work.

In section 3, we present the histories version of a classical scalar field theory: in
particular, we show how two Poincar\'{e} groups arise as an analogue of the two types
of time transformation in the non-relativistic history theory.

The free quantum scalar field theory is presented in section 4. We show that due to
the histories temporal structure previously introduced in \cite{Sav99}, manifest
Poincar\'{e} invariance is possible. Specifically, we show how different
representations of the history algebra---corresponding to different choices of
foliation---are realised on the {\em same\/} Fock space (notwithstanding the fact that
the different representations are unitarily inequivalent), and we show that they are
related in a certain way with Poincar\'{e} transformations.

\section{The History Projection Operator Approach}
The History Projection Operator (the, so-called, `HPO' approach) theory was a
development \cite{HPOHis} (emphasizing quantum {\em temporal\/} logic) of the
consistent-histories approach to quantum theory inaugurated by Griffiths, Omn\'es,
Gell-Mann and Hartle \cite{CoHis}. However, the novel temporal structure introduced in
\cite{Sav99} led to a departure from the original ideas on decoherence. In particular,
in our approach, emphasis is placed on the distinction between (i) the temporal logic
structure of the theory; and (ii) the dynamics \cite{SA00}.

In consistent-histories theory, a history is defined as a sequence of time-ordered
propositions about properties of a physical system, each of which can be represented,
as usual, by a projection operator. In normal quantum theory, it is not possible to
assign a probability measure to the set of all histories. However, when a certain
`decoherence condition' is satisfied by a set of histories, the elements of this set
{\em can\/} be given probabilities.

The probability information of the theory is encoded in the decoherence functional: a
complex function of pairs of histories which---in the original approach of Griffiths
{\em et al\/}---can be written as
\begin{equation}
d(\alpha,\beta)= {\rm tr}(\tilde C_\alpha^\dagger\rho \tilde
C_\beta) \label{decfun1}
\end{equation}
where $\rho$ is the initial density-matrix, and where the {\em class operator\/}
$\tilde C_\alpha$ is defined in terms of the standard Schr\"odinger-picture projection
operators $\alpha_{t_i}$ as
\begin{equation}
 \tilde C_\alpha:=U(t_0,t_1)\alpha_{t_1} U(t_1,t_2)
    \alpha_{t_2}\ldots U(t_{n-1},t_n)\alpha_{t_n}U(t_n,t_0)
\end{equation}
where $U(t,t')=e^{-i(t-t')H/\hbar}$ is the unitary time-evolution operator from time
$t$ to $t'$. Each projection operator $\alpha_{t_i}$ represents a proposition about
the system at time $t_i$, and the class operator $\tilde C_\alpha$ represents the
composite history proposition ``$\alpha_{t_1}$ is true at time $t_1$, and then
$\alpha_{t_2}$ is true at time $t_2$, and then \ldots, and then $\alpha_{t_n}$ is true
at time $t_n$''.

Isham and Linden developed the consistent-histories formalism further, concentrating
on its {\em temporal\/} quantum logic structure \cite{HPOHis}. They showed that
propositions about the histories of a system could be represented by {\em projection
operators\/} on a new, `history' Hilbert space. In particular, the history proposition
``$\alpha_{t_1}$ is true at time $t_1$, and then $\alpha_{t_2}$ is true at time $t_2$,
and then \ldots, and then $\alpha_{t_n}$ is true at time $t_n$'' is represented by the
{\em tensor product\/} $\alpha_{t_1}\otimes\alpha_{t_2}\otimes
\cdots\otimes\alpha_{t_n}$ which, unlike $\tilde C_\alpha$, {\em is\/} a genuine
projection operator, that is defined on the tensor product of copies of the standard
Hilbert space ${\cal H}_{t_1}\otimes{\cal H}_{t_2}\otimes\cdots\otimes{\cal H}_{t_n}$.
Hence the `History Projection Operator' formalism extends to multiple times, the
quantum logic of single-time quantum theory.

\paragraph*{The history space.} An important way of understanding the history Hilbert
space ${\mathcal F}$ is in terms of the representations of the `history group'---in
elementary systems this is the history analogue of the canonical group \cite{HPOHis}.
For example, for the simple case of a point particle moving on a line, the Lie algebra
of the history group for a {\em continuous\/} time parameter $t$ is described by the
history commutation relations
\begin{eqnarray}
{[}x_t,x_{t'}]&=&0  \label{HA1} \\
 {[}p_t,p_{t'}]&=&0   \label{HA2} \\
{[}x_t,p_{t'} ] &=& i \hbar\delta(t-t') \label{HA3}
  \end{eqnarray}
where $-\infty\leq t,\,t'\leq\infty$. It is important to note that these operators are
in the {\em Schr\"odinger\/} picture, and that the history algebra is invariant under
translations of the time index of these operators.

The choice of the Dirac delta-function in the right hand side of Eq.\ (\ref{HA3}) is
associated with the requirement that time be treated as a continuous variable. One
important consequence is the fact that the observables cannot be defined at sharp
moments of time but rather appear naturally as \textit{time-averaged}.

A unique representation of this algebra can be found by requiring the existence of an
operator analogue of a time-averaged Hamiltonian $H = \int_{-\infty}^{\infty}\! dt\,
H_t$, where $H_t$ is the standard Hamiltonian defined at a moment of time $t$
\cite{A60}.

\paragraph*{The Action and Liouville operators.}
One of the original problems in the development of the HPO theory was the lack of a
clear notion of time evolution, in the sense that, there was no natural way to express
the time translations from one time slot---that refers to one copy of the Hilbert
space ${\cal H}_t$---to another one, that refers to another copy ${\cal
H}_{t^{\prime}}$. The situation changed with the introduction of the {\em `action'
operator S\/}.

Indeed, the crucial step for constructing the temporal structure of the theory was the
definition in \cite{Sav99} of the action operator $S$---a quantum analogue of the
Hamilton-Jacobi functional \cite{Dirac}, written as
\begin{equation}
 S_{\kappa}:= \int^{+\infty}_{-\infty}\!\!dt\;( p_{t}\dot{x}_{t}- \kappa(t)H_{t}) ,
 \label{Def:op_S}
\end{equation}
where $\kappa(t)$ is an appropriate test function.

The first term of the action operator $S_{\kappa}$  Eq.\ (\ref{Def:op_S}) is identical
to the kinematical part of the classical phase space action functional. This
`Liouville' operator is formally written as
\begin{equation}
 V:= \int^{\infty}_{-\infty}\! dt\,(p_{t}\dot{x_{t}})  \label{liou}
\end{equation}
so that
\begin{equation}
 S_{\kappa} = V - H_{\kappa}.
\end{equation}

\subsection{The temporal structure}
A fundamental property of the HPO form of history theory is that the Liouville
operator $V$ and the Hamiltonian operator $H_{\kappa}$ generate two distinct types of
time transformation. The Liouville operator $V$ relates the Schr\"odinger-picture
operators associated with different time-$t$ labels, whereas $H_t$ is associated with
internal dynamical changes at the fixed time $t$ (with an analogous statement for the
smeared operator $H_\kappa$). The action operator $S_{\kappa}$ is thus the generator
of both types of time translation \cite{Sav99}.

More precisely, it was shown that there exist {\em two\/} distinct types of time
transformation. One---generated by the Liouville operator $V$---refers to time as it
appears in temporal logic, and it is related to $t$-label in Eqs.\
(\ref{HA1}--\ref{HA3}). The other---generated by the Hamiltonian---refers to time as
it appears in the implementation of dynamical laws, and it is related to the label $s$
in the `history Heisenberg picture' operator, that is hence defined in accord to the
novel conceptual issues introduced with the `two modes of time'
\begin{equation}
    x_t(s):=e^{isH/\hbar}\;x_t \;e^{-isH/\hbar}. \label{heis}
\end{equation}
where $H$ is defined to be $H_\kappa$ with $\kappa$ set equal to
$1$.

We will use the notation $x_{f}(s)$ for these history Heisenberg-picture operators
smeared with respect to the time label $t$, and we notice from Eq.\ (\ref{heis}) that
these quantities behave like standard Heisenberg-picture operators with a time
parameter $s$.

For any specific physical system these two transformations are intertwined with the
aid of the action operator $S$ as
\begin{equation}
e^{i\tau S/\hbar}\;x_f(s) \;e^{-i\tau S/\hbar}= x_{
f_\tau}(s+\tau)
\end{equation}
where $ f_\tau(t) := f(t+\tau)$, and where $S$ means $S_\kappa$ with $\kappa=1$.

\subsubsection*{Classical histories theory}
The continuous-time histories description has a natural analogue for classical
histories \cite{SA00}. In this scheme, the basic mathematical entity is the space $\Pi
= C(\mathR, \Gamma )$ of differentiable paths taking their value in the manifold
$\Gamma $ of classical states. Hence an element of $\Pi$ is a smooth path
$\gamma:\mathR \rightarrow \Gamma$. In effect, we associate a copy of the classical
state space with each moment of time, and employ differentiable sections of the
ensuing bundle over $\mathR$.

The key idea in this approach to classical histories is contained in the symplectic
structure on this space of temporal paths $\Pi$. For example, for a particle moving in
one dimension (with configuration coordinate $x$ and momentum coordinate $p$), the
history space $\Pi$ is equipped with a symplectic form
\begin{equation}
\omega = \int \!dt\, dp_t \wedge dx_t
\end{equation}
which generates the history Poisson brackets
\begin{eqnarray}
\mbox{\boldmath$\{$} x_t,x _{t^{\prime}}\mbox{\boldmath$\}\,$ }
&=& 0     \label{PB1} \\
\mbox{\boldmath$\{$} p_t,p
_{t^{\prime}}\mbox{\boldmath$\}\,$ } &=& 0 \label{PB2}\\
\mbox{\boldmath$\{$} x_t , p_{t^{\prime}}\mbox{\boldmath$\}\,$ }
&=& \delta(t-t^{\prime}) \label{PB3}
\end{eqnarray}
In general, given a function $f$ on $\Gamma$ we can define an
associated family $t\mapsto F_t$ of functions on $\Pi$ as
\begin{equation}
F_t(\gamma) := f(\gamma(t)) .          \label{Ff}
\end{equation}

In this way, all transformations implemented through the Poisson bracket in the normal
canonical theory, correspond to transformations in the history theory that {\em
preserve the time label $t$\/}. Indeed, for two families of functions $t\mapsto F_t$
and $t \mapsto G_t$ defined through (\ref{Ff}) we have
\begin{equation}
\{F_t,G_{t'} \} = L_t \delta(t,t'),
\end{equation}
where $L_t$ corresponds to the function $l$ on $\Gamma$
\begin{equation}
l = \{ f,g \}_{\Gamma}.
\end{equation}
In this way all relevant structures of the canonical theory can be naturally
transferred to the histories framework \cite{SA00}.

The Liouville, Hamilton and action functionals on $\Pi$ are defined respectively as
\begin{eqnarray}
    V ( \gamma )&:=& \int_{-\infty}^{\infty}\!dt \,{[p_t\dot{x_t}](\gamma)}    \\
 H ( \gamma )&:=& \int_{-\infty}^{\infty}\!dt \,{[H_t(p_t,x_t)](\gamma)}    \\
   S( \gamma ) &:=& V( \gamma )- H( \gamma )
\end{eqnarray}
where $\dot{x_t}(\gamma) = (\partial x_t / \partial t) (\gamma)$ is the velocity at
the time point $t$ of the path $\gamma$. These definitions are crucial for the
dynamics of the theory. In particular, $V$ and $H$ are the classical analogues of the
generators of the two types of time transformation in the history quantum theory
\cite{Sav99}.

The crucial result of classical histories theory is that one may deduce the equations
of motion in the following way \cite{Sav99}: a classical history ${\gamma}_{cl}$ is
the realised path of the system---{\em i.e.\/} a solution of the equations of motion
of the system---if it satisfies the equations
\begin{eqnarray}
 \{x_t , V\}(\gamma_{cl}) = \{x_t , H\}(\gamma_{cl}) \label{Ham1} \\
 \{p_t , V\}(\gamma_{cl}) = \{p_t , H\}(\gamma_{cl}) \label{Ham2}
\end{eqnarray}
where ${\gamma}_{cl}$ is the path $t\mapsto(x_t({\gamma}_{cl}) , p_t({\gamma}_{cl})
)$, and $x_t({\gamma}_{cl})$ is the position coordinate of the realised path
${\gamma}_{cl}$ at the time point $t$.

The above equations\ (\ref{Ham1}--\ref{Ham2}) are the history equivalent of the
canonical equations of motion. In particular, the symplectic transformation generated
by the history action functional $S(\gamma)$ leaves invariant the paths that are
classical solutions of the system:
\begin{eqnarray}
     \{x_t , S \}(\gamma_{cl})&=& 0  \label{clsol1} \\
     \{p_t , S \}(\gamma_{cl})&=& 0.  \label{clsol2}
\end{eqnarray}
More generally, any function $F$ on $\Pi$ satisfies the equation
\begin{equation}
     \{ F, S\} (\gamma_{cl})= 0 \label{action princ}.
\end{equation}

This is the way in which equations of motion appear in the classical history theory.
Notice that the role of the action as the generator of time transformations emerges
naturally in this classical case. Furthermore, the condition (\ref{action princ})
above emphasises the role of the Hamiltonian and Liouville functionals in histories
theory as generators of different types of  time transformation. It also clarifies the
new temporal structure that arises in history theory when compared with the standard
classical theory.

This result is of particular importance in the case of parameterised systems, where
the notion of time is recovered {\em after the phase space reduction \/} \cite{SA00}.

\section{Classical Scalar Field Theory}
\subsection{Background}
\subsubsection*{Standard canonical treatment}
In the Hamiltonian description of a free scalar field $\phi$ with mass $\tilde{m}$ on
Minkowski spacetime, the first step is to choose a spacelike foliation, which can be
specified by its normal---a unit time-like vector $n^{\mu}$.  We shall take the
signature of the Minkowski metric $\eta^{\mu\nu}$ to be $(+,-,-,-)$.

The first step is to select a specific foliation, and to choose a reference leaf
$\Sigma \simeq \mathR^3$ that is characterised by $t=0$, where $t$ is the natural time
label associated with the foliation.

The corresponding configuration space is the space $C^{\infty}(\Sigma)$ of all smooth
scalar functions $\phi(x)$ on $\Sigma$, while the phase space $\Gamma$  is its
cotangent bundle $T^*C^{\infty}(\Sigma)$ defined in an appropriate way\footnote{To
make these statements mathematically rigorous it would be necessary to invoke the
differential geometry of infinite-dimensional spaces like $C^\infty (\Sigma)$.
However, we do not need to become involved in such complexities here: for our purposes
it suffices to postulate the basic Poisson algebra relations
(\ref{CJIPB1}--\ref{CJIPB3}) that follow.}. The key point about this structure is that
the state space of fields is equipped with the Poisson brackets
\begin{eqnarray}
\{\phi(\underline{x}), \phi(\underline{x}') \} &=& 0  \\
 \{ \pi(\underline{x}), \pi(\underline{x}') \} &=& 0 \\
\{\phi(\underline{x}), \pi(\underline{x}') \} &=&
\delta(\underline{x}-\underline{x}').
\end{eqnarray}

\paragraph*{Poincar\'{e} group symmetry.} The relativistic scalar field  theory is
covariant under the action of the Poincar\'{e} group \cite{Itzy}. For a free massive
scalar field, the generators of time-translations $P^0$, space translations $P^i$,
spatial rotations $J^i$ and Lorentz boosts $K^i$ are respectively \footnote{They are
obtained by the use of Noether's theorem on the Lagrangian theory, and a Legendre
transform.}
\begin{eqnarray}
H &=& P^0 = \frac{1}{2} \int \!d^3\underline{x} \,[\pi^2 +
\partial_i \phi
\partial_i \phi + \tilde{m}^2
\phi^2] \\
P^i &=& \int \!d^3\underline{x}\, \pi \partial^i \phi \\
J^i &=& \frac{1}{2} \epsilon^{ijk} \int d^3\underline{x}\, \pi x_j \partial_k \phi \\
K^i &=& M^{0i} = \int \! d^3\underline{x} \,[t \pi \partial^i
\phi - x^i \frac{1}{2}(\pi^2 +
\partial_j \phi \partial_j \phi + \tilde{m}^2 \phi^2)] \label{Kboost}
\end{eqnarray}
where we note that the sub/superscripts $i,j,k$ refer to coordinates in the surface
$\Sigma$ that is spatial with respect to the chosen foliation vector $n$. Similarly,
the integrals above are all defined over $\Sigma$.

If we define the partial differential operator
\begin{equation}
(\Gamma f)(\underline{x}):=\left[(\eta^{\mu\nu}-n^\mu n^\nu)
\partial_\mu\partial_\nu +\tilde{m}^2 \right]f(\underline{x}), \label{Def:Kn}
\end{equation}
we can write the convenient expressions for the Hamiltonian and the boosts generator
as
\begin{eqnarray}
H &=& \frac{1}{2} \int \!d^3\underline{x} \,[\pi^2 + \phi \Gamma \phi] \\
K^i &=& \int \! d^3\underline{x} \,[t \pi \partial^i \phi -
\frac{1}{2} x^i ( \pi^2 + \phi \Gamma \phi)].
\end{eqnarray}

\bigskip

\subsection{Histories description for the classical scalar field}
In the histories formalism of a scalar field, the space of phase-space
histories\footnote {One may write a history version of the Lagrangian treatment,
however this description is not relevant to the immediate aims of this work.} $\Pi$ is
an appropriate subset of  the continuous Cartesian product $\times_t \Gamma_t $ of
copies of the standard state space $\Gamma$, each labeled by the time parameter $t$.
The choice of $\Gamma$ depends on the choice of a foliation vector $n^{\mu}$, hence
the space of histories also has an implicit dependence on $n^{\mu}$ and should
therefore be written as $^n \Pi$. Furthermore, we write $\Sigma_t = (n,t)$, the
space-like surface $\Sigma$ defined with respect to its normal vector $n$, and labeled
by the parameter $t$.

To be more precise, for each space-like surface $\Sigma_t$ we consider the state space
$\Gamma_t = T^* C^\infty (\Sigma_t)$. Then we define the fiber bundle with basis
$\mathR$ and fiber $\Gamma_t$, at each $t\in \mathR$. Histories are defined as the
cross-sections of the ensuing bundle, and the history space $^n\Pi$ is the space of
all smooth cross-sections of this bundle.

The Poisson algebra relations of the history theory are
 \begin{eqnarray}
\{\,\, \phi(X)\,, \phi(X^{\prime})\,\,\} &=& 0   \label{CJIPB1}\\
\{\,\, \pi(X)\,, \pi(X^{\prime}) \,\,\} &=& 0    \label{CJIPB2} \\
\{\,\, \phi(X)\,, \pi(X^{\prime}) \,\,\} &=& \delta^4 (X-
X^{\prime})\label{CJIPB3}
\end{eqnarray}
where $X$ and $X^{\prime}$ are space-time points. The field $\phi(X)$ and its
conjugate momentum $\pi(X^{\prime})$ are implicitly defined with respect to the
foliation vector $n^{\mu}$.

The definitions of the action $S$, Liouville $V$ and `Hamiltonian' $H$ functionals are
\begin{eqnarray}
S&:=& V - \frac{1}{2} \int \! d^4 \,X \,\,\{ \pi^2 (X) +
 \phi (X)\,\, {}^n\!\Gamma \,\phi (X)\}  \\
 V&:=& \int \! d^4 X \, \pi(X)n^{\mu} \partial_{\mu}\, \phi(X)  \\
 H&:=& \frac{1}{2} \int \!d^4 X \,\,\{ \pi^2 (X) +
 \phi (X)\,\, {}^n\!\Gamma \,\,\phi (X)\}
\end{eqnarray}
respectively, where again there is an implicit $n$ label on these three quantities;
and where $\Gamma$ is the differential operator
\begin{equation}
\Gamma(X):=\left[(\eta^{\mu\nu}-n^\mu n^\nu)
\partial_\mu\partial_\nu + {\tilde{m}}^2 \right] \label{Def:Gamman2}
\end{equation}
introduced above.

As we explained earlier, the variation of $S\[\gamma \]$ leaves invariant the paths
$\gamma_{cl}$ that are classical solutions of the system:
\begin{eqnarray}
     \{\phi(X) , S \}(\gamma_{cl})&=& 0  \label{clft1} \\
     \{\pi(X) , S \}(\gamma_{cl})&=& 0  \label{clft2}
\end{eqnarray}
As we shall now see, $H$ is the generator to the time averaged internal Poincar\'{e}
group.

\subsection{Poincar\'{e} symmetry}
The Poincar\'{e} group is the group of isometries of the Minkowski metric. Hence, any
field theory in Minkowski space-time needs to be covariant under the action of the
Poincar\'{e} group. As we shall now see, in a history theory---because of its
augmented temporal structure---the associated group theory leads to a particular
interesting result: namely, there are {\em two\/} distinct Poincar\'{e} groups that
act on the history space.

\subsubsection{The internal Poincar\'{e} group}
One significant feature of histories theory is that it gives a representation of the
temporal logic of the system that is {\em independent\/} of the dynamics involved.
Hence, propositions about the state of the system at different times are represented
by appropriate subsets of the space of paths. In the context of symmetries, however,
the temporal logic structure entails the following.

For each copy $\Gamma_t$ of the standard state space, there exists a Poincar\`{e}
group symmetry of the type one would expect in a canonical treatment of relativistic
field theory. On the other hand, in the history theory the state space is
heuristically the Cartesian product of such copies, and all physical quantities in the
standard treatment now appear as naturally time-averaged \cite{Sav99}. Hence one may
write time-averaged generators of the {\em internal } Poincar\'{e} groups, in a
covariant-like notation as
\begin{eqnarray}
\hspace*{-1cm}H \hspace*{-0.3cm}&=& \hspace*{-0.3cm}\frac{1}{2}\int \!d^4X \, \{
\pi(X)^2 + \phi(X)\, {}^n\!\Gamma \,\phi(X)\} \\
\hspace*{-1cm}P(m) \hspace*{-0.3cm}&=&\hspace*{-0.3cm} m_{\mu}\int \! d^4X \,\pi(X)\,
\partial^{\mu} \,\phi(X)  \label{Pmcl}  \\
\hspace*{-1cm}J(m) \hspace*{-0.3cm}&=& \frac{1}{2}n_{\mu} m_{\nu} \epsilon^{\mu
\nu \rho \sigma} \int \! d^4X \,\pi(X)\, X_{\rho} \,\partial_{\sigma}\, \phi(X)
\label{Jmcl} \\
\hspace*{-1cm}K(m)\hspace*{-0.3cm}&=& \hspace*{-0.3cm}\!m_{\mu} \int\! d^4X \, \{
n\!\cdot\! X \,\pi(X)\partial^{\mu}\phi(X) - \frac{1}{2} X^{\mu}[ \pi(X)^2 +
\phi(X)\,{}^n\!\Gamma \phi(X)\} \label{Kmcl}
\end{eqnarray}
where $m^\mu$ is an `$n$-spacelike' vector, {\em i.e.} one such
that $ n \cdot m:=n^\mu m^\nu\eta_{\mu\nu} = 0$.

Of special interest are the groups of canonical transformations generated by the
Hamiltonian generator $H$  and the boosts generator $K$. Note that a space-time point
$X$ can be associated with the pair $(t,\underline{x})\in\mathR\times\mathR^3 $, as
$X=tn + x_n$, where the three-vector $\underline{x}$ has been associated with a
corresponding $n$-spatial four-vector $x_n$  (i.e., $ n\cdot x_n=0 $); note that
$t=n\cdot X$. Then we define the classical analogue of the Heisenberg picture fields
as
\begin{eqnarray}
\phi(X)
\begin{array}{c}
  H \\
  \longrightarrow\\
\end{array}
\phi(X,s)
\end{eqnarray}
or
\begin{eqnarray}
\phi(t,\underline{x})
\begin{array}{c}
  H \\
  \longrightarrow\\
\end{array}
\phi(t,\underline{x},s): = \cos({}^n\Gamma^{\frac{1}{2}}s)
\phi(X) + \frac{1}{{}^n\Gamma^{\frac12}} \sin(
{}^n\Gamma^{\frac{1}{2}}s) \pi(X),
\end{eqnarray}
where $\phi(X):=\phi(t, \underline{x})$ and $\phi(X , s):= \phi(t
\,, \underline{x}\,, s)$. The square-root operator
${^n\Gamma}^{\frac{1}{2}}$, and functions thereof, can be defined
rigorously using the spectral theory of the self-adjoint, partial
differential operator ${^n\Gamma}$ on the Hilbert space
$L^2(\mathR^4,d^4X)$.

Notice also that the time label $t$ is not affected by this transformation since
$[n\cdot \partial, {}^n\!\Gamma] = 0$. For a fixed value of time $t$, the field
$\phi(t\,,\underline{x}\,,s )$ is the `Heisenberg-picture' field of the standard
canonical treatment.

The action of boost transformations is best shown upon objects $\phi(X,s) = \phi(t \,,
\underline{x}\,, s)$ as
\begin{equation}
\phi(t,\underline{x},s) \rightarrow \phi(t, \underline{x}',s'),
\end{equation}
where $(\underline{x}', s')$ and $(\underline{x},s)$ are related by the Lorentz boost
parametrised by $m^{\mu}$ as
\begin{eqnarray}
s' &=& \cosh |m| s + \frac{\sinh |m|}{|m|} x^i m_i \nonumber \\
{x^i}' &=& (\delta_{ij} - \frac{m^i m_j}{|m|^2}) x^j + \frac{m^i m_j}{|m|^2} \cosh|m|
x^j + \frac{\sinh|m|}{|m|} m^i s    \label{Lorentz}
\end{eqnarray}
where, as above, $x_i$ is the spatial part of $X$ with respect to $n$, so that $X= tn
+ x_n$ and $i=1,2,3$.

Hence, for each copy of the standard classical state space, there exists an `internal'
Poincar\`{e} group that acts on the copy of standard canonical field theory that is
labeled by the same $t$-time label.

\subsubsection{The external Poincar\'{e} group}
For each fixed $n$, there also exists an `{\em external}' Poincar\'{e} group with
generators
\begin{eqnarray}
\tilde{P}^{\mu} &=& \int \! d^4X \,\pi(X) \partial^{\mu} \phi(X) \\
\tilde{M}^{\mu \nu} &=& \int\! d^4X \,\pi(X) (X^{\mu}
\partial^{\nu} - X^{\nu}
\partial^{\mu}) \phi(X)
\end{eqnarray}
where $\mu, \nu = 0, 1, 2, 3$ and $\tilde{P}^{\mu}$ generate spacetime translations.
The $n$-spatial parts of the tensor $\tilde{M}^{\mu \nu}$ generate spatial rotations;
the time parts generates boosts.

The space translations and rotations are identical to those of the {\em internal\/}
Poincar\'{e} group. However the time translation and the boosts differ. Indeed, under
$V:= \tilde{P}^0$ we have
\begin{eqnarray}
\phi(t,\underline{x})
\begin{array}{c}
  V \\
  \longrightarrow\\
\end{array}
\phi(t+\tau ,\underline{x}) \\
\pi(t,\underline{x})
\begin{array}{c}
  V \\
  \longrightarrow\\
\end{array}
\pi(t+\tau ,\underline{x}).
\end{eqnarray}
where $\tau$ is the time translation generated by $V$. Thus, what we have shown here
is that the time-translation generator for the `external' Poincar\'{e} group is the
Liouville functional $V$.On the other hand, the boost generator $\tilde{K}^i =
\tilde{M}^{0i} $ generates Lorentz transformations of the type
\begin{eqnarray}
\phi(X) \rightarrow \phi(\Lambda X) \\
\pi(X) \rightarrow \pi(\Lambda X)
\end{eqnarray}
where for future convenience we write as  $\Lambda$  the element of the Lorentz group
obtained by exponentiation of the boost parameterised by  $m^i$.

Furthermore, under the action of this external group, the generators of the {\em
internal\/} Poincar\'{e} group transform as follows
\begin{eqnarray}
^{n}\!\!H &&
\begin{array}{c}
  \tilde{K} \\
  \longrightarrow \\
\end{array}
 \;\;\;^{\Lambda n }\!\!H \\
^{n}\!\!K(m)&&
\begin{array}{c}
  \tilde{K} \\
  \longrightarrow\\
\end{array}
\;\;\;^{\Lambda n}\!\!K (\Lambda m).
\end{eqnarray}
where we have now attached the explicit $n$ labels that were implicit in our previous
notation for these quantities. The action functional transforms in the same way
\begin{equation}
^nS \;\;\;\rightarrow \;\;\;^{\Lambda n}\!S
\end{equation}

Note that the action of the two groups  coincides on classical solutions
$\gamma_{cl}$:
\begin{eqnarray}
\{ \phi(X), K(m) \} (\gamma_{cl}) &=&  \{ \phi(X), \tilde{K}(m) \} (\gamma_{cl}) \\
\{ \pi(X), K(m) \}(\gamma_{cl}) &=& \{ \pi(X), \tilde{K}(m) \}
(\gamma_{cl})
\end{eqnarray}

We must emphasise again that the definition of $\Pi$ depends on the foliation vector.
Hence, so will the action of the Poincar\'{e} group. Here we deal with the scalar
field, for which this  dependence is not explicit. However, this dependence, and
analogue of the Poincar\'{e} group action is a major feature in systems where there is
an explicit foliation dependence. For example, this is the case  of general relativity
which is discussed in \cite{S01b}.

\bigskip
\bigskip

\section{Histories Quantum Scalar Field Theory}
\subsection{Background}
\subsubsection*{Canonical quantum field theory}
Canonical quantisation proceeds by looking for a representation of the {\em canonical
commutation relations}
\begin{eqnarray}
[\,\, \hat{\phi}(\underline{x})\,, \hat{\phi}(\underline{x}^{\prime})\,\,] &=& 0    \\
\[\,\, \hat{\pi}(\underline{x})\,, \hat{\pi}(\underline{x}^{\prime}) \,\,\] &=& 0   \\
\[\,\, \hat{\phi}(\underline{x})\,, \hat{\pi}(\underline{x}^{\prime}) \,\,\] &=& i\hbar
\delta^3 (\underline{x} -\underline{x}^{\prime})
\end{eqnarray}
on a Hilbert space which, in practice, is selected by requiring the existence of the
Hamiltonian as a genuine (essentially) self-adjoint operator.

For a free field, such a representation can be found on the Fock space $ {\mathcal{F}}
= \exp { L^2 ( {\mathR}^3\!, \,{d^3}\underline{x} ) } $ on which the fields can be
written in terms of the creation and annihilation operators $b$ and $b^{\dagger}$ that
define $\mathcal{F}$
\begin{eqnarray}
\hat{\phi}(\underline{x}) &=& \frac{1}{\sqrt{2}}
{}^n{\Gamma}^{-1/4}(\hat{b}(\underline{x}) +
\hat{b}^{\dagger}(\underline{x})) \\
\hat{\pi}(\underline{x}) &=& \frac{1}{\sqrt{2}} {}^n{\Gamma}^{1/4}
(\hat{b}(\underline{x}) - \hat{b}^{\dagger}(\underline{x}))
\end{eqnarray}
where
\begin {equation}
 [ b(\underline{x}), b^{\dagger}(\underline{x}^{\prime})] = \delta^{3} (\underline{x}-
\underline{x}^{\prime}) \label{bbstnd}.
\end{equation}

The (normal-ordered) Hamiltonian then reads\footnote{In momentum space we write $b$
and $b^{\dagger}$ from the well known relation $b_{{\bf k}} = \sqrt{\omega_{{\bf
k}}/2} \phi_{\bf k} + i /\sqrt{2 \omega_{{\bf k}}} \pi_{{\bf k}}$.}
\begin{equation}
\hat{H} = \int\!d^3\,\underline{x} \;\hat{b}^{\dagger}(\underline{x}) \,{}^n\Gamma \,
\hat{b}(\underline{x}) = \sum_{{\bf k}} \omega_{{\bf k}} \,\hat{b}^{\dagger}_{{\bf k}}
\,\hat{b}_{{\bf k}} \label{hatH}.
\end{equation}

\paragraph*{Poincar\'{e} group symmetry.}
A representation of the full Poincar\'{e} group exists on this Hilbert space. The
starting point is the generators of the classical theory, suitably normal-ordered to
correspond to well-defined operators. Substituting the fields in terms of creation and
annihilation operators, the generators can be written as
\begin{eqnarray}
\hat{P}^i &=& i \int\! d^3\underline{x}
\;\hat{b}^{\dagger}(\underline{x})
\,\partial^i \,\hat{b}(\underline{x})    \label{P^i} \\
\hat{J}^i &=& i \epsilon^{ijk} \int \!d^3\underline{x} \;
\,\hat{b}^{\dagger}(\underline{x})\, x_j\,
\partial_k \,\hat{b}(\underline{x})  \label{J^i} \\
\hat{K}^i &=& \int \!d^3\underline{x} \;
\hat{b}^{\dagger}(\underline{x})\, {}^n\Gamma^{1/4}\, x^i\;
{}^n\Gamma ^{1/4} \,\hat{b}(\underline{x}) \label{K^i}
\end{eqnarray}
These generators, together with $\hat{H}$ defined in Eq.\ (\ref{hatH}), satisfy the
Lie algebra relations of the Poincar\'{e} group.

In the canonical picture, the covariant fields are obtained by the Heisenberg
equations of motion
\begin{eqnarray}
\hspace*{-0.5cm}\hat{\phi}(\underline{x},s)\! &:=&
\!\!\!e^{\frac{i}{\hbar}s\hat{H}}\hat{\phi}(\underline{x})
e^{-\frac{i}{\hbar} s\hat{H}}\!\! = \cos({{}^n
\Gamma}^{\frac{1}{2}} s) \hat{\phi}(\underline{x}) +
{{}^n\Gamma}^{\frac{-1}{2}} \sin(
{{}^n \Gamma}^{\frac{1}{2}} s) \hat{\pi}(\underline{x})
\\[4pt]
\vspace{5cm} \hspace*{-0.5cm}\hat{\pi}(\underline{x},s)\!\!&:=&
\!\!\!e^{\frac{i}{\hbar} s\hat{H}}\hat{\pi}(\underline{x}) e^{-\frac{i}{\hbar} s
\hat{H}} \!= -{{}^n \Gamma}^{\frac{1}{2}}\sin( {{}^n \Gamma}^{\frac{1}{2}}
s)\hat{\phi}(\underline{x}) +
 \cos({{}^n\Gamma}^{\frac{1}{2}} s) \hat{\pi}(\underline{x})
\end{eqnarray}

The explicit automorphisms generated by the boosts may easily be calculated for the
Heisenberg picture creation and annihilation operators
\begin{equation}
\hat{b}(\underline{x},s) := e^{\frac{i}{\hbar} s\hat{H}} \; \hat{b}(\underline{x}) \;
e^{-\frac{i}{\hbar} s \hat{H}} = e^{- i s \, ^n \!\Gamma}
 \hat{b}(\underline{x})
\end{equation}
and they give
\begin{eqnarray}
e^{i m_i \hat{K}^i} \;\hat{b}(\underline{x},s) \;e^{-i m_i
\hat{K}^i} = \hat{b} (\underline{x}',s'), \label{canboost}
\end{eqnarray}
where the transformation $(\underline{x},s) \mapsto (\underline{x}^{\prime},
s^{\prime})$ is given by Eq.\ (\ref{Lorentz}), so that we can write
\begin{equation}
e^{i m_i \hat{K}^i} \;\hat{b}(\underline{x},s) \; e^{-i m_i
\hat{K}^i} = \hat{b}(\Lambda (\underline{x},s)).
\end{equation}
From this, one can write the explicit transformation laws for the Heisenberg fields
$\hat{\phi}(\underline{x},s)$ and $\hat{\pi}(\underline{x},s)$.

\paragraph*{Some questions that arise in the canonical treatment.} The first question
that arises in this standard treatment is whether the Poincar\'{e} transformations are
associated with any changes of foliation. Working canonically {\em there is no trace
of the foliation vector} on the Fock space defined by Eq.\ (\ref{bbstnd}), so this
question cannot readily be answered.

Being able to talk about foliations is a necessary step if we are to elucidate the
spacetime character of a quantum theory, in which the parameter $s$ of the Heisenberg
picture objects corresponds to the foliation time parameter in spacetime. For example,
the physical meaning of the parameter $s$ of the Heisenberg objects depends on the
choice of foliation vector.

\subsection{Histories Quantum Field Theory}
\paragraph*{Quantum mechanics histories.}
As we have already mentioned in section 2, the introduction of the history group
\cite{HPOHis} as an analogue of the canonical group relates the spectral projectors of
the generators of its Lie algebra with propositions about history phase space
quantities. This algebra is infinite-dimensional and therefore there exist infinitely
many representations. However the physically appropriate representation of the smeared
history algebra can be uniquely selected by the requirement that the time-averaged
energy exists as a proper self-adjoint operator \cite{HPOHis}. The resulting Hilbert
space has a natural interpretation as a continuous-tensor product: hence by this means
we also gain a natural mathematical implementation of the concept of `continuous'
temporal logic.

\subsubsection{Histories Hamiltonian algebra}
We shall now apply the histories ideas to relativistic quantum field theory on
Minkowski space-time. The representation of the history algebra is to be selected by
requiring that the time-averaged energy $H_{\chi} = \int\! d^4X \,\chi(t)\,H_t $,
(which is associated with history propositions about temporal averages of the energy)
exists as a proper essentially self-adjoint operator \cite{HPOHis}. In what follows,
for the sake of typographical simplicity we will no longer use hats to indicate
quantum operators.

We start with the abstract algebra
\begin{eqnarray}
[\,\, \phi(X)\,, \phi(X^{\prime})\,\,] &=& 0     \label{PhiChi1} \\
\[\,\, \pi(X)\,, \pi(X^{\prime}) \,\,\] &=& 0     \label{PhiChi2} \\
\[\,\, \phi(X)\,, \pi(X^{\prime}) \,\,\] &=& i \hbar \delta^4  \label{PhiChi3}
(X - X^{\prime})
\end{eqnarray}
where $X$ and $X^\prime$ are spacetime points.

In order to find suitable representations of this algebra we start with the Fock space
$ {\mathcal{F}}:= \exp{ {L}^2 ( {\mathR}^4, d^4 X ) } $ in which there is a natural
definition of creation and annihilation operators $b(X)$ and $b^{\dagger}(X)$ that
satisfy the commutation relations
\begin{eqnarray}
&&{[\,}b(X)\,, b( X^{\prime})\,] = 0                 \\
&&{[\,}b^\dagger(X)\,, b^\dagger(X^{\prime})\,] = 0     \\
&&{[\,}b(X)\,, b^\dagger(X^{\prime})] = \hbar\delta^4( X -
X^{\prime} ).
\end{eqnarray}

An appropriate representation of the Poincar\'{e} group can be defined by requiring
\begin{eqnarray}
 U(\Lambda)\,b(X)\,U(\Lambda) ^{\dagger} &=& b(\Lambda X) \\
  U(\Lambda)\mid \! 0\rangle &=& \mid \!0\rangle
\end{eqnarray}
where $\mid \! 0\rangle$ is the cyclic 'vacuum' state for the theory. Then clearly
history fields can be defined by
\begin{eqnarray}
 \phi(X) &:=& {1\over\sqrt2}\Big( b(X)+ b^{\dagger}(X)\Big) \label{phib}\\
 \pi(X) &:=& {1\over i\sqrt2}\Big( b(X)- b^{\dagger}(X)\Big). \label{pib}
\end{eqnarray}
and satisfy Eqs.\ (\ref{PhiChi1}--\ref{PhiChi3}). They also transform in the obvious
covariant way under the operators $U(\Lambda)$ introduced above.

It should be emphasized that the fields $\phi(X)$ and $\pi(X)$ thus defined {\em do
not\/} have any foliation vector dependence. However, an operator $H_{\chi}$ of the
time-averaged energy of the system {\em cannot\/} be well defined so that it depends
functionally on these fields in the usual way.

Hence we must seek a different, and more physically appropriate representation, for
the history algebra on the history Hilbert space $\mathcal{F}$.

We start by making a {\em fixed\/} choice of a unit time-like vector $n$ which we use
to foliate the four-dimensional Minkowski space-time.  It is clear that the
average-energy operator is itself dependent upon the choice of foliation $n$, and
therefore this must also be true for the elements of the history algebra. Hence to
emphasise that the physically appropriate representation depends on $n$ we rewrite the
history commutation relations as
\begin{eqnarray}
[\,\, ^n\!\phi(X)\,, ^n\!\phi(X^{\prime})\,\,] &=& 0  \label{hnqft1} \\
\[\,\, ^n\!\pi(X)\,, ^n\!\pi(X^{\prime}) \,\,\] &=& 0  \label{hnqft2} \\
\[\,\, ^n\!\phi(X)\,, ^n\!\pi(X^{\prime}) \,\,\] &=& i\hbar\delta^4 (X - X^{\prime})
 \label{hnqft3}
\end{eqnarray}
where $X$ and $X^\prime$ are spacetime points. The dependence of the representation of
the history algebra on the choice of the time-like foliation vector $n$ is indicated
by the upper left symbol for the field $^n\phi(X)$ and its `conjugate' $^n\pi(X)$.

\bigskip

One may also write the canonical version of the history algebra. Notice that---as in
the discussion above of classical history theory---in relating Eqs.\
(\ref{hnqft1})--(\ref{hnqft3}) with the canonical version of the history algebra the
three-vector $ \underline{x} $ may be equated with a four-vector $ x_n $ that
satisfies $ n\cdot x_n=0 $ (the dot product is taken with respect to the Minkowski
metric $ \eta_{\mu\nu}$ ) so that the pair $(t,\underline{x})\in\mathR\times\mathR^3 $
is associated with the spacetime point $ X=tn+x_n $ (in particular, $ t=n\cdot X $).
The canonical history commutation relations can be written therefore as
\begin{eqnarray}
 {[\,} ^n\!\phi( t \,,\underline{x} ), ^n\!\phi( t^\prime \,, \underline{x}^\prime
 )\,]&=&0 \label{1Dphiphi}\\
 {[\,} ^n\!\pi( t\,,\underline{x} ), ^n\!\pi( t^\prime \,,\underline{x}^\prime
 )\,]&=&0 \label{1Dpipi}\\
 {[\,} ^n\!\phi( t\,, \underline{x} ), ^n\!\pi( t^\prime \,, \underline{x}^\prime
 )\,]&=&i\hbar \delta ( t-t^\prime )\delta^3( \underline{x} -\underline{x}^\prime ),
 \label{1Dphipi}
\end{eqnarray}
where, for each $t\in\mathR$, the fields $ ^n\!\phi( t\,, \underline{x})$ and $
^n\!\pi( t\,, \underline{x})$ are associated with the spacelike hypersurface $\Sigma_t
=  (n,t)$, characterised by the normal vector $n$ and by the foliation parameter $t$.
In particular, the three-vector $\underline{x}$ in $ ^n\!\phi( t\,, \underline{x})$ or
in $ ^n\!\pi( t\,,\underline{x})$ denotes a vector in this space.

\bigskip

A central feature of the approach that is followed in this work for the histories
quantum field theory, is that  for all foliation vectors $n$, the corresponding
foliation-dependent representations of the history algebra Eqs.\
(\ref{hnqft1})--(\ref{hnqft3}) can all be realised on the {\em same\/} Fock space $
{\mathcal{F}} = \exp { L^2 ( \mathR^4, d^4 X ) } $ that also carries the `covariant'
fields $\phi(X)$ and $\pi(X)$ defined in  Eqs.\ (\ref{phib})--(\ref{pib}).

The foliation-dependent fields $^n\!\phi(X)$ and $^n\!\pi(X)$ are expressed in terms
of the covariant creation and annihilation operators of $ \exp { L^2 ( {\mathR}^4, d^4
X ) }$, and the related covariant fields $\phi (X)$ and $\pi (X)$ of Eqs.\
(\ref{hnqft1}--\ref{hnqft3}), as
\begin{eqnarray}
 ^n\!\phi(X) &=& -{1\over\sqrt2} {^n\Gamma}^{1/4} \Big( b(X)+
  b^{\dagger}(X)\Big) = {-^n\Gamma}^{1/4}\;\phi(X) \\ \label{nphib}
 ^n\!\pi(X) &=& {1\over i\sqrt2} {^n\Gamma}^{1/4} \Big( b(X)-
 b^{\dagger}(X)\Big) = {^n\Gamma}^{1/4}\;\pi(X), \label{npib}
\end{eqnarray}
and conversely,
\begin{eqnarray}
\hspace{-1cm}b(X)\!\!&=& {1\over\sqrt2}\Big(\phi(X)+i \pi(X)\Big) =
{1\over\sqrt2}\Big( {^n\!\Gamma}^{1/4}\,{}^n\!\phi(X)+ i \,{^n\!\Gamma}^{-
1/4}\,{}^n\!\pi(X)\Big)\label{Def:bnX} \\
\hspace{-1cm}b^{\dagger}(X)\!\!&=& {1\over
\sqrt2}\Big(\phi(X)-i \pi(X)\Big)= {1\over \sqrt2}\Big(
{^n\!\Gamma}^{1/4}\;{}^n\!\phi(X)- i \,{^n\!\Gamma}^{-1/4}\;{}^n\!\pi(X)\Big)
\label{Def:bdagnX}
\end{eqnarray}
where $^n\Gamma$ denotes the partial differential operator defined in Eq.\
(\ref{Def:Gamman2}) on the Hilbert space $ L^2 ( {\mathR}^4, d^4 X) $.

For a fixed foliation vector $n$, we seek a family of `internal' Hamiltonians $ ^n
H_{t}$, $t\in\mathR$, whose explicit formal expression may be deduced from the
standard quantum field theory expression to be
\begin{equation}
\hspace{-0.1cm}^n\!H_t:={1\over 2}\!\int \!\!d^4X\left\{
{^n\!\pi(X)}^2\!+\! (n^\mu n^\nu\!-\!\eta^{\mu\nu})\:\partial_\mu
\:^n\!\phi(X)\:\partial_\nu \:^n\! \phi(X)\! + \!\tilde{m}^{2}\:
{^n\!\phi(X)}^2\right\}\delta(t\!-\!n\!\cdot \!\! X)
\label{Def:Hnt}
\end{equation}
The corresponding smeared expression (which must be normal-ordered to be well-defined)
is
\begin{eqnarray}
\hspace{-0.1cm}^n\!H_{\chi}\!\!\!\!\!&:=&\int_{-\infty}^\infty
\!\!dt\;\chi(t)\,^n\!H_t \label{Def:Hnchi} \\
&=&\hspace{-0.3cm}{1\over 2}\! : \!\int \!\! d^4X\left\{{^n\!\pi(X)}^2\!\!\!+
\!\!(n^\mu n^\nu\!\!-\!\!\eta^{\mu\nu})\;\partial_\mu \;^n\!\phi(X)\;\partial_\nu\;
^n\!\phi(X) + \tilde{m}^2 \; {^n\!\phi(X)}^2\right\}\chi(n\!\cdot\!X):    \nonumber
\end{eqnarray}
where $\chi$ is a real-valued test function.

We next augment the history algebra with the following commutation relations that
would be satisfied by the operators ${}^n\!H(\chi)$, if they existed,
\begin{eqnarray}
{[\,} ^n\!H_{\chi},\, ^n\!\phi(X)\,] &=& -i\hbar\chi(n\cdot X)\;
^n\!\pi(X)
 \label{[HchiphiX]}\\
{[\,} ^n\!H_{\chi},\, ^n\!\pi(X)\,] &=& i\hbar\chi(n\cdot X)\;
^n\!\Gamma\;
 ^n\!\phi(X) \label{[HchipiX]}\\
 {[\;\,} ^n\!H_{\chi} ,\, ^n\!H_{\chi^{\prime}}\;\,] &=& 0.
\end{eqnarray}

If the operators ${}^n\!H$ existed, the above commutation relations would give rise to
the transformations
\begin{eqnarray}
e^{\frac{i}{\hbar} ^n\!H_{\chi}}\;
^n\!\phi(X)\; e^{\frac{-i}{\hbar}^n\!H_{\chi}}\!\!\!&=& \label{AutphiX}\\
&=& \cos\left[\chi(n\!\cdot
\!X){^n\!\Gamma}^{\frac{1}{2}}\right]\, ^n\!\phi(X)\!\! +
{^n\!\Gamma}^{\frac{-1}{2}} \sin\left[\chi(n\!\cdot
\!X) {^n\!\Gamma}^{\frac{1}{2}}\right]\, ^n\!\pi(X)   \nonumber \\
e^{\frac{i}{\hbar} ^n\!H_{\chi}}\;
^n\!\pi(X)\; e^{\frac{-i}{\hbar}^n\!H_{\chi} }\!\!\!&=& \label{AutpiX}\\
&=& -{^n\!\Gamma}^{\frac{1}{2}}\sin\left[\chi(n\!\cdot
\!X){^n\!\Gamma}^{\frac{1}{2}}\right]\, ^n\!\phi(X) +
\cos\left[\chi(n\!\cdot \!X){^n\!\Gamma}^{\frac{1}{2}}\right]\,
^n\!\pi(X)  \nonumber
\end{eqnarray}
Note that the expression $\chi(n\!\cdot \!X){^n\Gamma}^{\frac{1}{2}}$ is unambiguous
since, viewed as an operator on $L^2(\mathR^4,d^4X)$, multiplication by $\chi(n\!\cdot
\!X)$ commutes with ${^n\Gamma}^{\frac{1}{2}}$.

The right hand side of Eqs.\ (\ref{AutphiX})--(\ref{AutpiX}) defines an automorphism
of the history algebra Eqs.\ (\ref{hnqft1})--(\ref{hnqft3}), and all that remains is
to show that these automorphisms are unitarily implementable in this representation.
To this end, we use Eqs.\ (\ref{Def:bnX})--(\ref{Def:bdagnX}) to prove that
\begin{equation}
    e^{i \,^n\!H_{\chi}/\hbar}\,b(X)\,e^{-i \,^n\!H_{\chi}/\hbar}=
        e^{-i\,\chi(n\cdot X) \;{^n\Gamma}^{\frac{1}{2}}}\,b(X).   \label{Autb(X)}
\end{equation}
However, the operator defined on $L^2(\mathR^4, d^X)$ by
\begin{equation}
    (O(\chi)\psi)(X):=e^{-i\chi(n\cdot X) \;{^n\Gamma}^{\frac{1}{2}}}\psi(X)
\end{equation}
is easily seen to be unitary, and hence we conclude \cite{HPOHis} that the desired
quantities $^nH_{\chi}$ exist as self-adjoint operators on the Fock space $\cal F$
associated with the creation and annihilation operators $b^\dagger(X)$ and $b(X)$. The
spectral projectors of these operators $^nH_{\chi}$ represent propositions about the
time-averaged value of the energy in the spacetime foliation determined by $n$.

To conclude: for each {\em fixed\/} choice of a foliation vector $n$, we have a
physically meaningful representation of the history algebra Eqs.\
(\ref{hnqft1})--(\ref{hnqft3}) on the Hilbert space $ {\mathcal{F}} = \exp { L^2 (
{\mathR}^4, d^4 X ) } $. Thus the same Hilbert space $ \mathcal{F}$ carries {\em
all\/} different representations---for different choices of $n$---of the quantum field
theory history algebra.

\subsubsection{The action operator}
We now define the action $^{n}\!S_{\chi}$ and the Liouville $^{n}\!V$ operators as
normal-ordered versions of their classical analogues
\begin{eqnarray}
 ^{n}\!S_{\chi}&=& ^{n}\!V - \frac{1}{2} : \int\! d^4 X \,\,\{ ^{n}\!\!\pi^2
(X)+ ^{n}\!\!\phi (X)\,\, ^{n}\!\Gamma \,\,^{n}\!\!\phi (X)\} \chi(n\cdot X) : \\
 ^{n}\!V&=& : \int_{-\infty}^\infty \!\!d^4 X \, ^n\!\pi(X)\, n^{\mu} \partial_{\mu}\,
 ^n\!\phi(X) :
\end{eqnarray}
The automorphisms of the history algebra that are generated by the action and
Liouville operators are
\begin{eqnarray}
 e^{i s ^n\!S_{\chi}/\hbar}\,b(X)\,e^{-i s ^n\!S_{\chi}/\hbar}&=&
 e^{-i\int_{s^{\prime} - s}^{s^{\prime} + s} ds^{\prime}\, \chi(nX+ s^{\prime})
 {^n\!\Gamma}^{\frac{1}{2}} - s n^{\mu} {\partial}_{\mu}}\,b(X)\label{AutoS(X)}
 \\[4pt]
 e^{i s ^n\!V/\hbar}\,b(X)\,e^{-i s ^n\!V/\hbar}&=&
 e^{- s \,n^{\mu} {\partial}_{\mu}}\,b(X),    \label{AutoV(X)}
\end{eqnarray}
and are easily shown to be unitarily implementable. In what
follows, the real-valued smearing function $\chi$ is set equal to
$\chi(t)=1$ for every $t \in \mathR$.

\subsection{Poincar\'{e} group covariance}
A significant feature of the histories formalism is the temporal structure of the
theory. It introduces a new approach to the concept of time, in which time is
distinguished as an ordering parameter (logical structure), and as an evolution
parameter (dynamics). In particular, as we have already shown in non-relativistic
quantum mechanics \cite{Sav99}, the Liouville operator $^nV$ generates time
translations with respect to the `external' $t$-time parameter, and the Hamiltonian
operator $^nH$ generates time translations with respect to the `internal' evolution
$s$-time parameter. The action operator $^nS$ generates {\em both\/} types of time
transformations; it is the time generator for the histories theory for solutions of
the equations of motion \footnote{In histories theory the physical time translation
generator is the action operator $^nS$; both Liouville $^nV$ and Hamiltonian $^nH$
operators are time translation generators that correspond to two different aspects
(two modes) of the notion of time. However, only $^nS$ is related to the actual
physical time parameter, in analogy with the standard theory where the Hamiltonian
$^nH$ is the time translation generator. }. The same construction is true for a
histories quantum field theory.

\bigskip

The invariance of standard quantum field theory under the Poincar\'{e} group, has been
a difficult issue to address for many years. In a canonical treatment of quantum field
theory, the Schr\"{o}dinger-picture fields depend on the reference frame ({\em i.e.},
choice of foliation). In order to demonstrate manifest independence of this choice
with the aid of Heisenberg-picture fields, one still has to contend with the
foliation-dependence of the Hamiltonian that generates the Heisenberg fields.

In histories theory, the enhanced temporal structure enables the study of a
Poincar\'{e} group transformation  between different foliations. In particular we will
show that different representations corresponding to different foliation vectors $n$,
are related by Lorentz boosts of the `external' Poincar\'{e} group:
\begin{equation}
 U(\Lambda)\,{}^{n}\!\phi(X)\,{U(\Lambda)}\!\!^{-1} = {}^{\Lambda n}
 \!\phi(\Lambda X)
\end{equation}
and where the time translations generator is closely related to
the `Liouville' operator $V$.

\paragraph*{The Heisenberg-picture operators.}
We first define the Heisenberg-picture analogue of the scalar field, to illustrate the
different time translations associated with the two time labels. We use a similar
notation to that in the classical case: {\em i.e.}, the Heisenberg-picture field is
written as $^n\!\phi(X,s) = {}^n\!\phi (t, \underline{x}, s)$, where the space-time
point $X = (t , \underline{x})$ is expressed in coordinates adapted to $n$. Thus
\begin{eqnarray}
\!\!^n\!\phi(X,s) &=& \,^{n}\!{\phi} ( t,\underline{x}, s) := e^{\frac{i}{\hbar}s\,
^n\!H}\;
{}^{n}\!\phi(t,\underline{x})\; e^{ -\frac{i}{\hbar}s\,^n\!H } \nonumber\\
\vspace*{2.5cm}&=& \cos \left(s\,{^n\!\Gamma}^{\frac{1}{2}}\right)\, ^n\!\phi(X)\! +
{^n\!\Gamma}^{\frac{-1}{2}} \sin \left(s\,
{^n\!\Gamma}^{\frac{1}{2}}\right)\, ^n\!\pi(X) \hspace{2cm} \label{phiheis}\\
\vspace*{2.5cm}\!\! ^n\!\pi(X,s) &=& \,^{n}\!{\pi} ( t,\underline{x} , s) :=
e^{\frac{i}{\hbar}s\,^n\!H}\;  {}^{n}\!\pi
( t,\underline{x})\; e^{ -\frac{i}{\hbar}s\,^n\!H} \nonumber \\
\vspace*{2.5cm}&=&  - \! {^n\!\Gamma}^{\frac{1}{2}} \sin \left(s\,
{^n\!\Gamma}^{\frac{1}{2}} \right)\, ^n\!\phi(X)\! + \cos  \left(s\,
{^n\!\Gamma}^{\frac{1}{2}} \right)\, ^n\!\pi(X) . \label{piheis}
\end{eqnarray}
The different types of time translation are particularly easy to see by studying the
action of the Liouville $^nV$ and action $^nS$ operators on the Heisenberg-picture
fields $b(X,s)$
\begin{eqnarray}
  e^{i\tau ^{n}\!\!H }\,  b (t,\underline{x} , s)\, e^{ -i\tau ^{n}\!\!H }  &:=&
  b (t\,,\underline{x}\,, s+\tau)                    \\ \label{bH}
  e^{i\tau ^{n}\!V}\,  b(t, \underline{x} , s)\, e^{ -i\tau ^{n}\!V }  &:=&
  b (t+ \tau \,,\underline{x} \,, s)                  \\ \label{bV}
   e^{i\tau ^{n}\!S}\,  b (t,\underline{x} , s)\, e^{ -i\tau ^{n}\!S }  &:=&
   b (t + \tau \,, \underline{x} \,, s + \tau)    \label{bS}
\end{eqnarray}
The label $s$ corresponds to the `internal' time of the unitary Hamiltonian time
evolution, while $t$ corresponds to the `external' time that labels the time-ordering
of events in a history for the Shr\"{o}dinger-picture operators.

\subsubsection{The internal Poincar\'{e} group}
As we showed previously, each fixed choice of foliation vector $n$ corresponds to a
\textit{different} representation of the history algebra on the \textit{same} Fock
space ${\mathcal{F}} = \exp{L^2({\mathR}^4,d^4X)}$. Hence, we may heuristically
say\footnote{The physical quantities in histories appear naturally space-time
averaged, therefore they are smeared with appropriate test functions. Strictly
speaking, quantities labeled at moments of time are not well-defined mathematically.},
that, for a given vector $n$, and for each value of the associated time $t$, there
will be a Hilbert space ${\mathcal{H}}_t$ that carries an independent copy of the
standard quantum field theory. In particular, there exists a representation of the
Poincar\'{e} group associated with each spacelike slice $(n\:,t)$, where $t\in
\mathR$.

In what follows, a particularly important role will be assigned to the averaged
`internal' Poincar\'{e} group. For example, we define the averaged energy ${}^{n}\!H
:= \int\! d^4X \, {}^n\!H_t$ that generates translations on the $s$-time parameter of
the Heisenberg-picture fields $^n\!\phi(X,s)= \:^n\!\phi(t\,,\underline{x}\,,s)$,
without affecting the `external' $t$-time parameter:
\begin{eqnarray}
^n\!\phi(X,s) =
\begin{array}{c}
  {}^{n}\!H \\
  \longrightarrow\\
\end{array}
\:^n\!\phi(X,s+s')   \\
 ^n\!\pi(X,s) =
\begin{array}{c}
  {}^{n}\!H \\
  \longrightarrow\\
\end{array}
\:^n\!\pi(X,s+s').
\end{eqnarray}

The expressions for the `internal' Poincar\'{e} generators of spatial translations
$P^i$, and rotations $J^i$ can be written in direct analogy with the expressions Eqs.\
(\ref{Pmcl})--(\ref{Kmcl}) of the classical case. We use the normal-ordered
expressions
\begin{eqnarray}
P (m) &=& : \int\! d^4X\, \pi(X)\, m^{\mu} \partial_{\mu} \,\phi(X) :  \nonumber\\
 &=& i \int\! d^4X \, b^{\dagger}(X) \, m^{\mu} \partial_{\mu} \, b(X) \label{Pint} \\
J (m) &=&  \frac{1}{2} n_{\mu} m_{\nu} \epsilon^{\mu \nu \rho
\sigma} : \int\!
d^4X \, \pi(X)\, X_{\rho} \,\partial_{\sigma}\, \phi(X) :  \nonumber \\
&=& i \frac{1}{2} n_{\mu} m_{\nu} \epsilon^{\mu \nu \rho \sigma}
\int\! d^4X \, b^{\dagger}(X) \, X_{\rho} \,\partial_{\sigma}\,
b(X) \label{Jint}
\end{eqnarray}
We have used an `pseudo-covariant' notation by employing a $n$-spacelike vector $m$
({\em i.e.,} such that ($n_{\mu} m^{\mu} = 0$). Note that the terms involving a pair
of creation operators, or a pair of annihilation operators, can be shown to vanish
through integration by parts.

Of particular interest, is the action of the boost generator $^nK
(m)$ defined as
\begin{eqnarray}
 \hspace*{-7cm}^n\!K(m)\!\! \!&=& \!m_{\mu} :\!\int \!d^4X \, [n\!\cdot X \, ^n\!\pi
 \,{\partial^{\mu}} \,^n\!\phi
 - \frac{1}{2} X^{\mu} \,( ^n\!\pi^2 + \,^n\!\phi \; ^n\!\Gamma \, ^n\!\phi)]: \\
\!\!\!&=& \!\int \!d^4X \, b^{\dagger}(X)\,
^{n}\!{\Gamma}^{\frac{1}{4}} \,X^{\mu} m_{\mu}\,\, ^{n}\!
{\Gamma}^{\frac{1}{4}} \,b(X).
\end{eqnarray}

The key feature of the boost generator $^n\!K(m)$ is that it mixes the
\textit{$s$}-time parameter with the three-vectors \underline{x}. The action of these
boost transformations is most clearly seen on the Heisenberg objects $\phi(X,s) = \phi
(t,\underline{x}, s)$
\begin{equation}
 ^{{\rm int}} U(\Lambda) \; ^n \!\phi ( t\,,\underline{x}\,, s) \; ^{{\rm int}}
 {{U(\Lambda)}\!\!^{-1}}= \,^n \!\phi (t\,,\Lambda(\underline{x}\, , s) ) ,
\end{equation}
where $^{{\rm int}} U({\Lambda}):= e^{iK(m)}$ is the unitary operator that generates
Lorentz transformations, and $\Lambda$ is the Lorentz transformation generated by $m$.

At this point we note the action of the internal Poincar\'{e} group on the action
$^n\!S$, Hamiltonian $^n\!H$ and Liouville $^n\!V$ operators respectively:
\begin{eqnarray}
^{{\rm int}} U({\Lambda})\,\, {}^{n}\!H\ {}^{{\rm int}}
 {U({\Lambda})}^{-1} = {}^{n}\!H   \\
^{{\rm int}} U({\Lambda})\,\, {}^{n}\! V\ {}^{{\rm int}}
 {U({\Lambda})}^{-1} = {}^{n }\!V  \\
^{{\rm int}} U({\Lambda})\,\, {}^{n}\!S\ {}^{{\rm int}}
 {U({\Lambda})}^{-1} = {}^{n}\!S .
\end{eqnarray}
As we would expect from standard canonical quantum field theory, we see that the above
operators remain invariant under the `internal' Lorentz transformations.

\subsubsection{External Poincar\'{e} group}
A key result in histories classical field theory is that there also exists a
second---the `external'---Poincar\'{e} group symmetry of the theory, with generators
\begin{eqnarray}
\tilde{P}^{\mu} &=& : \int\! d^4X \, \pi(X) \,\partial^{\mu} \,\phi(X) : \\
\tilde{M}^{\mu \nu} &=& : \int\! d^4X \, \pi(X) \, (X^{\mu}
\partial^{\nu} - X^{\nu}
\partial^{\mu}) \, \phi(X) :
\end{eqnarray}
Note that these definitions use the covariant fields $\phi(X)$ and $\pi(X)$ that
satisfy the algebra Eqs.\ (\ref{PhiChi1})--(\ref{PhiChi3}) rather than the
foliation-dependent fields $^n\!\phi(X)$ and $^n\!\pi(X)$ of Eqs.\
(\ref{hnqft1})--(\ref{hnqft3}). However, many of the generators of the external
Poincar\'{e} group are exactly the same whether one uses covariant fields expressions
or foliation-dependent ones: they differ only for the case of the boosts generators
${}^nK(m)$.

In particular, the Liouville operator $\tilde{P}^0 = V$, given by the expression
\begin{eqnarray}
V &=& : \int \! d^4 X\, \pi(X)n^\mu \partial_{\mu}\, \phi(X):\\
\nonumber &=& i \int \! d^4 X\, b^{\dagger}(X) n^{\mu}
\partial_{\mu} b(X)
\end{eqnarray}
generates translations on the time label $t$.

The space translations and rotation generators are identical to those of the internal
Poincar\'{e} group Eqs.\ (\ref{Pint}--\ref{Jint}). However the external boost
generator $\tilde{K}(m)$ differs from the internal one ${}^nK(m)$, and hence it is of
particular interest to study the action of the former.

The generator of time-translations $V$ acts on Schr\"odinger picture objects as
\begin{eqnarray}
 ^{n}\!\!\phi(X) = ^{n}\!\!\phi(t\,,\underline{x})
\begin{array}{c}
  V \\
  \longrightarrow\\
\end{array}
  \:^{n}\!\!\phi(t+\tau\,,\underline{x}) \\
  ^{n}\!\!\pi(X) = ^{n}\!\!\pi(t\,,\underline{x})
\begin{array}{c}
  V \\
  \longrightarrow\\
\end{array}
 \:\:^{n}\!\!\pi(t+\tau \,,\underline{x}).
\end{eqnarray}

The `external' boost generator $\tilde{K}(m)$ is
\begin{eqnarray}
\tilde{K}(m) &=& : \: \int_{-\infty}^{\infty}\! d^4X \:
\pi (X)\, T_m \,\phi (X) \: :\\
 &=&  i \int_{-\infty}^{\infty}\! d^4X \:  b^{\dagger} (X)\, T_m \, b (X)
\end{eqnarray}
where we define the operator $T_m$ as
\begin{equation}
(T_m f)(X):= n_{\mu}m_{\nu} (X^{\mu}{\partial}^{\nu} -
X^{\nu}{\partial}^{\mu}) f(X).
\end{equation}
and $n \cdot m = 0$. Then the boost generator $\tilde{K}(m)$ acts on the fields
$^{n}\!\phi(X)$ as
\begin{equation}
 ^{{\rm ext}} U(\Lambda)\,\, {}^{n}\!{\phi} (X)\ {}^{{\rm ext}}
 {U(\Lambda)}^{-1} = \;^{\Lambda n}\!{\phi} ({\Lambda}(X) ),
\end{equation}
and it mixes the $t$-time parameter with the three-vector \underline{x}. However, the
crucial point is that $\tilde{K}(m)$ generates Lorentz transformations {\em on the
foliation vector $n$} as well.

This can be viewed as a demonstration of explicit Poincar\'{e} covariance, as we can
see from the action of the external Lorentz transformations on the Heisenberg-picture
fields ${}^{n}\!{\phi} (X\,,s)$ as
\begin{equation}
 ^{{\rm ext}} U({\Lambda})\,\, {}^{n}\!{\phi} (X,s)\ {}^{{\rm ext}}
 {U({\Lambda})}^{-1} = \,^{\Lambda n}\!{\phi} ({\Lambda}(X,s) ).
\end{equation}
The generators of the {\em internal } Poincar\'e group transform under the action of
the {\em external \/} Poincar\'{e} group as
\begin{eqnarray}
{}^{{\rm ext}} U(\Lambda)\; {}^{n}\!H \ {}^{{\rm ext}}
 {U(\Lambda})^{-1} &=& \,{}^{\Lambda n}\!H \\
{}^{{\rm ext}} U(\Lambda)\; {}^{n}\!\tilde{K}(m)\; {}^{{\rm ext}}
 {U(\Lambda)}^{-1} &=& \,{}^{\Lambda n }\! \tilde{K}(\Lambda m).
\end{eqnarray}
Of considerable importance is the fact that the action operator ${}^n\!S$ transforms
in the same way:
\begin{equation}
^{{\rm ext}} U(\Lambda)\,\, {}^{n}\!S\ {}^{{\rm ext}}
 {U(\Lambda)}^{-1} = {}^{\Lambda n}\!S
\end{equation}
Hence the action of the external Poincar\'{e} group {\em relates\/} representations of
the theory that {\em differ \/} with respect to the foliation vector $n$. As we shall
see in the following section, this is crucial when we discuss the Poincar\'{e}
invariance of probabilities.

\bigskip

In summary, we have showed that the history version of quantum field theory carries
representations of two Poincar\'{e} groups. The `internal' Poincar\'{e} group is
defined in analogy to the one in the standard canonical treatment of the theory. It
corresponds to time-translations with respect to the `internal' $s$-time parameter of
histories theory.  The Lorentz part of the `external' Poincar\'{e} group intertwines
representations of the theory associated with different choices of foliation, all of
which however are realised on the {\em same\/} Fock space $\mathcal{F}$. It
corresponds to time-translations with respect to the `external' $t$-time parameter.

The translation parts of these two types of Poincar\'{e}
transformation---corresponding to the relations between the $t$ time parameter and
kinematics, and the $s$ time parameter and dynamics---have very significant analogues
in the case of the histories version of general relativity \cite{S01b}.

\subsubsection{The decoherence functional}
\paragraph*{`Classical' coherent states.}
In \cite{Sav99}, we showed how a classical-quantum relation can be nicely described in
histories theory by using the history analogue of coherent states. In the histories
formalism, a non-normalised coherent state vector is written as \cite{HPOHis}
\begin{equation}
 |\exp z\rangle = \oplus^{\infty}_{n=0}(n!)^{-\frac{1}{2}}
  (\otimes|z\rangle)^n.
\end{equation}
The corresponding normalised coherent states  can be obtained  by unitary
transformations of the vacuum state as
  \begin{equation}
     |z \rangle :=    \frac{1}{\sqrt{\langle
\exp z| \exp z \rangle}}  |\exp z \rangle = U[f,h] |0\rangle
  \end{equation}
where $U[f,h]$ is the Weyl operator defined as
  \begin{equation}
   U[f,h]:= e^{\frac{i}{\hbar}(\;^n\!\phi(f)\:-\:^n\!\pi(h)\;)} ,
  \end{equation}
and $f$ and $h$ are smearing functions that belong to $L^2({\mathR}^4,d^4X)$. We write
the normalised coherent state $|z \rangle$ corresponding to the pair $f,h$ as $|f,h
\rangle$. In this context we know that $f$ and $h$ correspond to classical values and
therefore correspond to a path on classical phase space. In this correspondence, the
functions $f$ and $h$ are the classical values of the field $\phi (X)$ and its
conjugate momenta $\pi (X)$, respectively.

The set of all  coherent states is  independent of the choice of foliation since these
coherent states are eigenstates of the annihilation operator $b (X)$, which is
foliation independent. However, the physical  identification of the vector  $|
z\rangle $ with  a phase space path {\em is\/} foliation-dependent since it depends on
the Weyl operator, which itself depends on the choice of the representation of the
history algebra  on the Fock space $\mathcal{F}$\footnote{Given a complex path $z$,
the classical phase space path $(f,h)$ is defined by the foliation-dependent
expression $z = {}^n\!\Gamma^{1/4} f + i \;{}^n \!\Gamma \!^{-1/4} h $}. One should
recall that the space of classical histories $\Pi$ is itself dependent on the choice
of foliation.

\bigskip

So far our discussion of the histories version of quantum field theory has been at the
level of field algebras and group transformations. However, in histories formalism
physically crucial `probabilistic' information is contained in the decoherence
functional.

In this HPO formalism, the  most general form for the decoherence functional of a pair
of history propositions $\alpha$, $\beta$ is
\begin{eqnarray}
d(\alpha, \beta) = Tr_{{\mathcal{F}} \times {\mathcal{F}}}
\left(\alpha \otimes \beta \,\Xi \right),
\end{eqnarray}
in terms of an operator $\Xi$ on  ${\mathcal{F}} \times {\mathcal{F}}$ \cite{ILS}.

In our case,  the operator $\Xi$ reads
\begin{equation}
 \Xi := \langle 0 | \rho_{-\infty}| 0 \rangle({\mathcal{S}}_{cts}
  {\mathcal{U}})^{\dagger}\otimes({\mathcal{S}}_{cts}{\mathcal{U}}),
\end{equation}
in terms of the operator ${\mathcal{S}}_{cts}{\mathcal{U}}$ that we proved in
\cite{Sav99} that it is an implicit function of the action operator: therefore there
is an implicit dependence of $\Xi$ on the foliation vector $n$. The matrix elements of
${\mathcal{S}}_{cts}{\mathcal{U}}$ in a coherent state basis can written in terms of
the classical action functional $S[f,h]$ as
\begin{equation}
\langle f,h|{\mathcal{S}}_{cts}{\mathcal{U}}|f,h \rangle = e^{i
S[f,h]}.
\end{equation}

The explicit relation of ${\mathcal{S}}_{cts}{\mathcal{U}}$ with the action operator
$^{n}\!S$ is as follows. For a general operator $A$ on $L^2({\mathR}^4 , d^4X)$ one
can define an operator $\Gamma(A)$ on ${\mathcal{F}}$ as
\begin{equation}
\Gamma(A) |\exp z \rangle = | Az \rangle.
\end{equation}
In our case we have
\begin{eqnarray}
e^{is {}^n\!S } = \Gamma ( e^{i s ^n\!\sigma} ) \\
  {\mathcal{S}}_{cts}{\mathcal{U}} = \Gamma ( 1 + i ^n\!\sigma ),
\end{eqnarray}
in terms of the operator $^{n}\!\sigma = n^{\mu} \partial_{\mu} - ^n\!\Gamma^{1/2}$.
Hence, the decoherence functional depends on the representation through the phase
space action $^nS$.

This raises the critical issue of the physical meaning of the fact that the formalism
appears to depend on a specific choice of the foliation vector $n$. We have seen above
that the representation of the phase space quantities by Hilbert space operators
depends on $n$, and that there exist unitary intertwiners between different
representations given by the boosts of the external Poincar\'e group. As has been
discussed in \cite{Stef}, a transformation law for the observables by means of a
unitary operator $U$
\begin{equation}
\alpha \rightarrow  \alpha' = U  \alpha U^{\dagger}
\end{equation}
implies that the operator $\Xi$ of the decoherence functional, carrying a label for
the foliation dependence $n$, ought to transform as
\begin{equation}
^n \Xi \rightarrow ^{\Lambda n} \!\Xi = (U \otimes U) \;^n \Xi
\;(U^{\dagger} \otimes U^{\dagger})
\end{equation}
so that the values of the decoherence functional (corresponding to probabilities and
correlation functions of the theory) {\em are representation-independent}
\begin{equation}
^{\Lambda n}\!d(\,^{\Lambda n}\!\alpha\,, ^{\Lambda n}\!\beta \,)
= {}^n\!d( ^n\!\alpha \,, ^n\!\beta),
\end{equation}
where $^{\Lambda n}\!d$ is the decoherence functional defined with reference to the
operator $^{\Lambda n}\!\Xi$.

In our case  we have  $U = e^{i \tilde{K}(m)} = {}^{ext}\!U(\Lambda)$. This changes
the foliation dependence of the fundamental fields $^n\!\phi(X)$ and  $^n\!\pi(X)$,
and hence of any observable ${}^n\!\alpha$ that depends upon them
\begin{equation}
{}^n\! \alpha  \rightarrow {}^{\Lambda n}\!\alpha : = U(\Lambda)\;
^n\!\alpha \;U(\Lambda)^{\dagger}
\end{equation}
Some physically interesting examples of observables, in this sense, are integrals
$\int\!d^X \, ^n\!\phi(X) f(X)$ of fields $^n\!\phi(X)$, smeared with appropriate test
functions $f(X)$, that satisfy $f(X)= f(\Lambda(X)$; another example is any space-time
average of the normal-ordered polynomial functions of these fields.

In order to see, how the boosts generator acts on ${}^n\Xi$, it suffices to check its
action on ${\mathcal{S}}_{cts}{\mathcal{U}}$. This is
\begin{equation}
U \;^n\!{ {\mathcal{S}}_{cts} {\mathcal{U}} } \; U^{\dagger} =
\Gamma ( 1 +i {e^{-T_m}}\; ^{n}\!\sigma \; e^{T_m} ) = \Gamma (1
+ i \,^{\Lambda n}\!\sigma)
\end{equation}
Consequently the operator $^n\!\Xi$ transforms as $ ^{n}\!\Xi \rightarrow ^{\Lambda
n}\!\Xi $. Hence the values of the decoherence functional are foliation independent
\begin{equation}
{}^n\!d({}^n\!\alpha \,,{}^n\!\beta) = {}^{\Lambda n}\!d(
{}^{\Lambda n}\!\alpha \,, {}^{\Lambda n}\!\beta).
\end{equation}

\section{Conclusions}
We have studied both the classical and the quantum history versions of scalar field
theory. We have showed that, in both cases, the crucial feature of the history field
theory is the appearance of two Poincar\'{e} groups, in direct analogy to the two
types of time transformation that characterizes the history formalism. The internal
Poincar\'{e} group is related to time as an ordering parameter (the Hamiltonian $H$ is
the time translations generator), and it is in analogy to the Poincar\'{e} group of
standard field theory. On the other hand, the external Poincar\'{e} group is related
to time as a parameter of evolution (the Liouville $V$ is the time translations
generator), and it is of particular interest for the quantum case, as it relates
representations of the quantum field theory, for different choices of foliation, with
Poincar\'{e} transformations.

These results will be proved of great importance in the study of history general
relativity theory in \cite{S01b}. In particular, the histories formalism is suitable
to deal with issues that lie at the level of the interplay between quantum theory and
the spacetime structure. The present work focuses on quantum field theory in a fixed
spacetime, however the techniques involved and the concepts introduced, have been able
to precisely identify the relation between the quantum mechanical observables and the
necessary notion of the spacetime foliation. Many issues are raised at the level of
the meaning of reference frames in quantum theory---a foliation corresponds to a
reference frame---and more importantly at the level of quantum gravity.

The latter is eventually the aim of the histories programme, and this involves a
further elucidation of the meaning of spacetime in a quantum theory. What strikes us
as relevant at present is that, one might have to disentangle between the two
different views of spacetime transformations: the {\em passive} and the {\em active}
view. This is subtly hinted by the fact that the transformations generated by the
external Poincar\'e group should be viewed in the passive sense, since the argument
$X$ cannot be identified with a fixed, absolute spacetime point {\em in all
representations}.

In order to successfully address the above issues we must first study the history
version of general relativity; this is the context of the forthcoming paper
\cite{S01b}.

\vspace{2cm}

\noindent{\large\bf Acknowledgements}

\noindent I would like to thank Charis Anastopoulos and Chris Isham for very helpful
discussions. I gratefully acknowledge support from the L.D. Rope Third Charitable
Settlement and from the EPSRC GR/R36572 grant.

\end{document}